\documentclass[aps,prb,longbibliography,groupedaddress,twocolumn,10pt]{revtex4-2}
\usepackage{graphicx}
\usepackage{mathtools}
\usepackage{amsmath}
\usepackage{amssymb}
\usepackage{bbm}
\usepackage{bm}
\usepackage{cancel}
\usepackage{xspace}
\usepackage{braket}
\usepackage{xcolor}
\usepackage{siunitx}
\usepackage{mathrsfs}
\usepackage{comment}
\usepackage{placeins}

\usepackage[export]{adjustbox}
\usepackage{wrapfig}
\usepackage{bbold}
\usepackage{booktabs} 
\usepackage[colorlinks=true]{hyperref} 

\newcommand{\ee}{\mathrm{e}}  
\DeclareMathOperator*{\ii}{i} 
\newcommand*\dd{\mathop{}\!\mathrm{d}}

\newcommand{\kel}[1]{\underline{#1}} 

\definecolor{hblue}{RGB}{0,80,255}
\definecolor{hred}{RGB}{255,80,0}

\begin{document}

\title{Extension of the iterated perturbation theory \\ at arbitrary fillings to nonequilibrium steady states}

\author{Tommaso Maria Mazzocchi}
\email[]{mazzocchi@tugraz.at}
\author{Enrico Arrigoni}
\email[]{arrigoni@tugraz.at}
\affiliation{Institute of Theoretical and Computational Physics, Graz University of Technology, 8010 Graz, Austria}

\date{\today}


\begin{abstract}
We extend the Kajueter-Kotliar [Phys. Rev. Lett. 77, 131 (1996)] iterated perturbation theory (KK-IPT) away from half filling to nonequilibrium steady states. We benchmark the resulting nonequilibrium KK-IPT approach against the auxiliary master equation approach (AMEA), whose accuracy is controlled in and out of equilibrium. As expected, in equilibrium, KK-IPT reproduces the AMEA results for different fillings with high accuracy at the level of both spectral properties and electron densities. Out of equilibrium, we study quantum transport across a correlated impurity and compute the differential conductance and spectral functions. We find very good agreement between nonequilibrium KK-IPT and AMEA in the parameter regime where the latter is reliable, in particular at moderate temperatures and biases. Although a controlled benchmark is not available in the low-temperature, low-bias regime, where AMEA becomes less reliable, we show that this nonequilibrium KK-IPT impurity solver satisfies the exact spectral sum rules for the first and second moments to high accuracy throughout the entire parameter range studied. These results support nonequilibrium KK-IPT as an approximate description of nonequilibrium steady states away from half filling. At the same time, comparing against AMEA the double occupancy obtained from the nonequilibrium Galitskii-Migdal expression for the interaction energy shows that the deviation from AMEA remains small near half filling for moderate and large values of the bias, but grows markedly away from half filling, delineating the regime in which the method can be trusted quantitatively rather than merely qualitatively.
\end{abstract}


\maketitle

\section{Introduction}\label{sec:intro}

The main challenge in dynamical mean-field theory (DMFT)~\cite{ge.ko.92,ge.ko.96} is the accurate and efficient solution of the quantum impurity problem arising from the mapping of a correlated lattice onto an effective Anderson impurity model (AIM). Within this framework, the effect of the surrounding lattice is encoded in a self-consistently determined hybridization function, the calculation of which requires repeated solutions of the AIM and therefore constitutes the main computational bottleneck of the method. Indeed, the overall performance of a DMFT calculation is largely determined by the availability of a fast and reliable impurity solver capable of handling the relevant interaction strengths, temperatures, and orbital complexities, a limitation that becomes particularly severe in nonequilibrium problems~\cite{ao.ts.14}. 

A variety of numerical techniques have been developed for this purpose, including quantum Monte Carlo (QMC)~\cite{gu.mi.11,we.co.06}, numerical renormalization group (NRG)~\cite{wils.75,bu.co.08}, exact diagonalization (ED)~\cite{ca.kr.94}, as well as strong-coupling~\cite{ke.ki.70,ec.we.10} methods, each of which provides complementary advantages and limitations. Continuous-time QMC solvers are numerically exact, but they may suffer from severe fermionic (dynamical) sign problems or become inefficient at very low temperatures or in nonequilibrium settings. In addition, the analytic continuation from imaginary to real frequencies is an ill-posed problem that typically broadens spectral features, particularly away from the Fermi level. By contrast, NRG methods yield highly accurate spectral information at low energies and temperatures, yet their spectral resolution typically deteriorates at higher energies. ED-based techniques work directly on the real-frequency axis and approximate the hybridization through a finite number of bath sites, at the cost of an exponential growth of the many-body Hilbert space with system size.  While plain (unitary) ED has the disadvantage that it cannot address a steady state, this drawback is resolved within the auxiliary master equation approach (AMEA)~\cite{do.nu.14,we.lo.23} by adopting the open quantum systems formalism~\cite{ma.ga.22,ga.ma.22,ma.we.23,ma.we.25}. The {\em inchworm} extension of the QMC algorithm can mitigate the sign problem~\cite{co.gu.15} and be employed to compute nonequilibrium properties of quantum systems within the Keldysh formalism~\cite{an.do.17}, including direct access to steady-state properties~\cite{er.gu.23,er.bl.24}. However, the treatment of long-time dynamics or steady-state regimes in strongly correlated impurity problems remains computationally demanding. Besides extensions of NRG~\cite{ande.08,han.25u} and the aforementioned AMEA, tensor-train and tensor-cross-interpolation-based approaches have lately emerged as promising impurity solvers~\cite{nu.je.22,ki.we.25}. This motivates the search for computationally inexpensive impurity solvers that can capture the essential correlation effects across a wide parameter regime in nonequilibrium DMFT. This is particularly true for steady-state calculations, where numerically controlled approaches are available but are often considerably more demanding than approximate schemes.

Among approximate impurity solvers, iterated perturbation theory (IPT) offers a useful combination of low computational cost and, in equilibrium, good agreement with exact diagonalization for both the AIM and the Hubbard model, especially at half filling~\cite{ge.kr.93}. Although perturbative in origin, IPT is constructed to recover both the weak-coupling and atomic limits and can therefore be viewed as an interpolation scheme between them, which helps explain its usefulness beyond the weak-coupling regime~\cite{ka.ko.96,le.pa.14}. While plain IPT is unreliable away from half filling, Kajueter and Kotliar~\cite{ka.ko.96} have introduced a correction to the electronic self-energy (SE) that generalizes it to arbitrary fillings, see also Refs.~\cite{ma.fl.82,ma.lo.86,po.we.97}. This approach has been subsequently refined to capture the Fermi-liquid behavior~\cite{ar.se.12}, to describe multiorbital correlated materials in equilibrium~\cite{da.mo.16} and to characterize the metal-insulator transition in the Bethe lattice within DMFT~\cite{vanl.22}. Extensions of IPT including higher order weak-coupling diagrams with and without self-consistency have been introduced to study nonequilibrium dynamics~\cite{ts.we.13}, although the extension by Kajeuter and Kotliar has not been implemented for nonequilibrium steady states.

Building on Ref.~\cite{ka.ko.96}, we extend the Kajueter-Kotliar iterated perturbation theory (KK-IPT) impurity solver for arbitrary filling to nonequilibrium steady-state conditions within the Keldysh Green's function formalism~\cite{schw.61,keld.65,ha.ja.98}. In particular, we investigate the quantum transport across a correlated impurity and benchmark the resulting nonequilibrium KK-IPT approach against AMEA. The latter provides a controlled reference for nonequilibrium steady-state problems at arbitrary fillings, even though it becomes less reliable in the low-temperature, low-bias regimes.
For this reason, we also compute the exact sum rules for the first and second spectral moment of the impurity as benchmarks to assess the range of validity of the proposed KK-IPT approach.

The remainder of this paper is organized as follows. In Sec.~\ref{sec:model}, we briefly introduce the model at hand. In Sec.~\ref{sec:method}, we present the nonequilibrium KK-IPT solver. Section~\ref{sec:results} is devoted to the discussion of the results, and Sec.~\ref{sec:conclusions} contains our concluding remarks together with a brief outlook on possible applications and extensions of the method.

\section{The Anderson impurity model}\label{sec:model}

The Hamiltonian of the AIM reads
\begin{equation}\label{eq:tot_hamiltonian}
\hat{H} = \hat{H}_{\text{imp}} + \hat{H}_{\text{bath}} + \hat{H}_{\text{hyb}},
\end{equation}
with the impurity Hamiltonian given by
\begin{equation}
\hat{H}_{\text{imp}} = \epsilon_{\text{f}} \sum_{\sigma} \hat{f}^{\dagger}_{\sigma} \hat{f}_{\sigma} + U \hat{n}_{\uparrow} \hat{n}_{\downarrow},
\end{equation}
where $\hat{f}^{\dagger}_{\sigma}$ ($\hat{f}_{\sigma}$) creates (annihilates) an electron of spin $\sigma$ on the impurity and $\hat{n}_{\sigma} \equiv \hat{f}^{\dagger}_{\sigma} \hat{f}_{\sigma}$ is the corresponding particle number operator. The onsite energy and onsite repulsion are denoted by $\epsilon_{\text{f}}$ and $U$, respectively~\footnote{Throughout this work the onsite energy is chosen to be spin independent.}. The bath Hamiltonian is given by
\begin{equation}
\hat{H}_{\text{bath}} = \sum_{\lambda p} \left( \epsilon_{\lambda} - \mu_{\lambda} \right) \hat{c}^{\dagger}_{\lambda p} \hat{c}_{\lambda p},
\end{equation}
where $\epsilon_{\lambda}$ and $\mu_{\lambda}$ are the onsite energy and chemical potential of the lead $\lambda = \left\{ \text{l}, \text{r} \right\}$ and $\hat{c}^{\dagger}_{\lambda p}$ ($\hat{c}_{\lambda p}$) creates (annihilates) an electron with quantum number $p$ on lead $\lambda$. The interaction between the impurity and the leads' degrees of freedom reads
\begin{equation}
\hat{H}_{\text{hyb}} = \sum_{\lambda p \sigma} t_{\lambda} \hat{c}^{\dagger}_{\lambda p} \hat{f}_{\sigma} + \text{H.c.},
\end{equation}
with $t_{\lambda}$ the hopping amplitude between the leads and the impurity. The hybridization function resulting from the coupling between the impurity and the leads is given by
\begin{equation}
  \underline{\Delta}(\omega) = \sum_{\lambda} t^{2}_{\lambda} \underline{L}_{\lambda}(\omega),
\end{equation}
where $\underline{L}_{\lambda}$ denotes the lead Green's function (GF). As usual, underlined symbols denote the Keldysh structure
\begin{equation}\label{eq:keld_struct}
\underline{X}(\omega) =
\begin{pmatrix}
X^{\text{R}}(\omega) & X^{\text{K}}(\omega) \\
0 & X^{\text{A}}(\omega)
\end{pmatrix},
\end{equation}
with $X^{\text{R}}$, $X^{\text{K}}$ and $X^{\text{A}} (= \left(X^{\text{R}}\right)^{\dagger})$ the retarded, Keldysh and advanced components of the GF. The greater ($>$) and lesser ($<$) components read
\begin{equation}\label{eq:lssgtr_from_RK}
X^{\gtrless}(\omega) = X^{\text{K}}(\omega)/2 \pm \ii \text{Im} X^{\text{R}}(\omega).
\end{equation}
If not stated otherwise, the retarded component $L^{\text{R}}_{\lambda}$ is chosen to be independent of the lead with
\begin{equation}\label{eq:leads_gf}
 \begin{split}
  \text{Im} L^{\text{R}}(\omega) & = - \bigl( 1 - f_{\text{\tiny FD}}(\omega+D; T_{\text{fict}}) \bigr) f_{\text{\tiny FD}}(\omega-D; T_{\text{fict}}), \\
  L^{\text{K}}_{\lambda}(\omega) & = 2 \ii \text{Im} L^{\text{R}}(\omega) \left( 1 - 2 f_{\text{\tiny FD}}(\omega-\mu_{\lambda}; T) \right),
 \end{split} 
\end{equation}
where $D$ defines the half bandwidth of the leads, $T_{\text{fict}}$ is a fictitious temperature that smears out the edges of the retarded component and $f_{\text{\tiny FD}}(\omega-\mu; T) = \left[ \ee^{\left( \omega - \mu \right)/T} + 1 \right]^{-1}$ the Fermi-Dirac distribution function. Throughout this work we use $\epsilon_{\lambda} = 0$, $T_{\text{fict}}=1/2$ and $t_{\lambda} = 1/\sqrt{2}$. To recover the equilibrium condition it is enough to set $\mu_{\text{l}} = \mu_{\text{r}} = \mu$; in this case, $\mu$ will then be set to zero for simplicity.

\section{Nonequilibrium iterated perturbation theory}\label{sec:method}

In this section, we briefly review KK-IPT for arbitrary fillings~\cite{ka.ko.96} and present its extension to nonequilibrium steady states, referred to here as nonequilibrium KK-IPT. The KK-IPT expands the electronic SE up to second order in the local interaction $U$ by combining three Weiss-field propagators $\underline{\mathscr{G}}_{0}$. For a given hybridization function $\underline{\Delta}$, the Keldysh components of the Weiss field $\underline{\mathscr{G}}_{0}$ read
\begin{equation}\label{eq:gen_weiss_field}
\begin{split}
\mathscr{G}^{\text{R}}_{0}(\omega) & = 1/\left( \omega + \mu_{0} - \Delta^{\text{R}}(\omega) \right), \\
\mathscr{G}^{\text{K}}_{0}(\omega) & = \Delta^{\text{K}}(\omega) \mid \mathscr{G}^{\text{R}}_{0}(\omega) \mid^{2},
\end{split}
\end{equation}
where $\mu_{0}$ is an auxiliary chemical potential to be determined self-consistently as it will be discussed below. Following Ref.~\cite{ka.ko.96}, we restrict ourselves here to the paramagnetic case, $n_{\uparrow} = n_{\downarrow} = n$. The extension of nonequilibrium KK-IPT to spin-dependent cases is formally straightforward, following Ref.~\cite{po.he.98}, but its numerical performance remains to be assessed and will be the subject of future work.

In this manuscript, we retain the KK-IPT Ansatz for the retarded component of the SE~\cite{ka.ko.96}. Its extension to the Keldysh component is then straightforward within a steady-state Keldysh framework~\cite{schw.61,keld.65,ha.ja.98}. The electronic SE~\cite{ka.ko.96} then reads
\begin{equation}\label{eq:extended_ipt_se_ansatz}
\begin{split}
\Sigma^{\text{R}}(\omega) & = n U + \left( 1 - \Sigma^{\text{R}}_{\text{\tiny B}}(\omega) \right)^{-1} \Sigma^{\text{R}}_{\text{\tiny A}}(\omega), \\
 \Sigma^{\text{K}}(\omega) & = \left( 1 - \Sigma^{\text{R}}_{\text{\tiny B}}(\omega) \right)^{-1} \times \\
& \left[ \Sigma^{\text{K}}_{\text{\tiny A}}(\omega) + \Sigma^{\text{K}}_{\text{\tiny B}}(\omega) \left(1 - {\Sigma^{\text{R}}_{\text{\tiny B}}}^{\ast}(\omega) \right)^{-1} {\Sigma^{\text{R}}_{\text{\tiny A}}}^{\ast}(\omega) \right],
\end{split}
\end{equation}
where we have introduced the shorthand notation
\begin{equation}\label{eq:shorthand_notation_coeffs}
\begin{split}
\Sigma^{\text{X}}_{\text{\tiny A}}(\omega) \equiv A \, \tilde{\Sigma}^{\text{X}}(\omega), \\
\Sigma^{\text{X}}_{\text{\tiny B}}(\omega) \equiv B \, \tilde{\Sigma}^{\text{X}}(\omega), \\
\end{split}
\end{equation}
with $\text{X} = \left\{ \text{R}, \text{K} \right\}$ and $^{\ast}$ denoting complex conjugation. The derivation of the Keldysh component of the SE in Eq.~\eqref{eq:extended_ipt_se_ansatz} can be found in Appedix~\ref{sec:keld_se_derivation}. The explicit form of the (real) coefficients $A$ and $B$ can be found in Ref.~\cite{ka.ko.96} and will be given below. It is worth noting that at half filling, for which \( A = 1 \), \( B = 0\) and $n=1/2$, we recover the particle-hole symmetric case, i.e. $\Sigma^{\text{R}}(\omega) = U / 2 + \tilde{\Sigma}^{\text{R}}(\omega)$ and $\Sigma^{\text{K}}(\omega) = \tilde{\Sigma}^{\text{K}}(\omega)$, in agreement with previous work~\cite{ma.ar.24}. The Keldysh components of the second-order diagram $\tilde{\Sigma}^{\text{R/K}}$ read
\begin{equation}\label{eq:se0_ret_kel}
\begin{split}
\text{Im} \tilde{\Sigma}^{\text{R}}(\omega) & = \frac{1}{2\ii} \left( \tilde{\Sigma}^{>}(\omega) - \tilde{\Sigma}^{<}(\omega) \right), \\
\tilde{\Sigma}^{\text{K}}(\omega) & = \tilde{\Sigma}^{>}(\omega) + \tilde{\Sigma}^{<}(\omega).
\end{split}
\end{equation}
The greater and lesser components in~\eqref{eq:se0_ret_kel} are given by
\begin{equation}\label{eq:lss_gtr_se}
\tilde{\Sigma}^{\gtrless}(\omega) = U^{2} \int \frac{\dd\omega_{2}}{2\pi} \ I^{\gtrless}(\omega+\omega_2) \mathscr{G}^{\lessgtr}_{0}(\omega_{2}),
\end{equation} 
where 
\begin{equation}\label{eq:lss_gtr_G0}
I^{\gtrless}(\omega+\omega_2) \equiv \int \frac{\dd\omega_{1}}{2\pi} \ \mathscr{G}^{\gtrless}_{0}(\omega + \omega_{2} - \omega_{1}) \mathscr{G}^{\gtrless}_{0}(\omega_{1}).
\end{equation}
As usual, the real part of the {\em retarded} SE in~\eqref{eq:se0_ret_kel} can be computed via the {\em Kramers-Kronig} relation
\begin{equation}
\text{Re} \tilde{\Sigma}^{\text{R}}(\omega) = \frac{1}{\pi} \int \dd \omega^{\prime} \frac{\text{Im} \tilde{\Sigma}^{\text{R}}(\omega)}{\omega^{\prime} - \omega}.
\end{equation}
The coefficients $A$ and $B$ in Eq.~\eqref{eq:shorthand_notation_coeffs} are given by~\cite{ka.ko.96}
\begin{equation}\label{eq:ext_ipt_coeffs}
\begin{split}
A & = \frac{n \left( 1 - n \right)}{n_{0}\left( 1 - n_{0} \right)}, \\
B & = \frac{\left( 1 - n \right) U + \epsilon_{\text{f}} + \mu_{0}}{n_{0}\left( 1 - n_{0} \right) U^{2}},
\end{split}
\end{equation}
where the particle number $n$ is obtained by
\begin{equation}\label{eq:occupation_num}
n = -\ii \int \frac{\dd\omega}{2\pi} G^{<}(\omega),
\end{equation}
and \( G^{<}(\omega) \) is computed as in Eq.~\eqref{eq:lssgtr_from_RK} from the retarded and Keldysh components of the impurity GF
\begin{equation}\label{eq:imp_gf}
\begin{split}
G^{\text{R}}(\omega) & = \frac{1}{{\mathscr{G}^{\text{R}}_{0}(\omega)}^{-1} - \mu_{0} - \epsilon_{\text{f}} - \Sigma^{\text{R}}(\omega)}, \\
G^{\text{K}}(\omega) & = \left( \frac{\mathscr{G}^{\text{K}}_{0}(\omega)}{\mid \mathscr{G}^{\text{R}}_{0}(\omega) \mid^{2}}+ \Sigma^{\text{K}}(\omega)\right) \mid G^{\text{R}}(\omega)  \mid^{2}.
\end{split}
\end{equation}
The quantity $n_{0}$ in Eqs.~\eqref{eq:ext_ipt_coeffs} is the particle number density calculated from the Weiss field,
\begin{equation}\label{eq:occupation_num0}
 n_{0} = -\ii \int \frac{\dd\omega}{2\pi} \mathscr{G}^{<}_{0}(\omega).
\end{equation}
We stress that the occupations~\eqref{eq:occupation_num} and~\eqref{eq:occupation_num0} are calculated without explicit use of the Fermi-Dirac distribution and thus are valid also in nonequilibrium conditions.

The algorithm employed in this work is based on the so-called IPT-$n_{0}$ approximation~\cite{po.he.98,da.mo.16}, in which the occupation of the Weiss field is constrained to equal that of the impurity. Despite its simplicity, this scheme performs well in equilibrium~\cite{po.he.98}. By contrast, other schemes, such as those enforcing the Luttinger theorem, are known to be less accurate as they fail to reproduce higher-order moments; see, e.g., the discussion in Ref.~\cite{po.he.98}. 

The workflow is as follows. For a given hybridization function $\underline{\Delta}(\omega)$, we start from an initial guess $\left( n^{\text{\tiny trial}}, \mu^{\text{\tiny trial}}_{0} \right)$ and compute the Weiss field, the SE, and the impurity GF according to Eqs.~\eqref{eq:gen_weiss_field}, \eqref{eq:extended_ipt_se_ansatz}, and \eqref{eq:imp_gf}, respectively. We then determine the self-consistent solution $(n,\mu_{0})$ by solving the two equations
\[
f_{1}\left[n^{\text{\tiny trial}}, \mu^{\text{\tiny trial}}_{0}\right]
\equiv
n^{\text{\tiny trial}} + \ii \int \frac{\dd\omega}{2\pi} G^{<}(\omega) = 0
\]
and
\[
f_{2}\left[n^{\text{\tiny trial}}, \mu^{\text{\tiny trial}}_{0}\right]
\equiv
n^{\text{\tiny trial}} + \ii \int \frac{\dd\omega}{2\pi} \mathscr{G}^{<}_{0}(\omega) = 0
\]
iteratively~\footnote{We note that, out of equilibrium, the Friedel sum rule discussed in previous work~\cite{ka.ko.96,da.mo.16} does not apply, and its proper generalization remains unclear. Nonetheless, benchmark comparisons with the AMEA impurity solver indicate that the IPT-$n_0$ approximation performs satisfactorily in all cases considered so far.}. As already reported in Ref.~\cite{ka.ko.96}, convergence is typically achieved within 10 iterations.~\footnote{Each impurity solver run discussed in this work was initialized from the \emph{same} fixed cold-start initial guess ($\mu^{\text{trial}}_0=0$, $n^{\text{trial}}=1/2$) with no continuation or warm-starting between neighboring bias points: each $(U,T,\epsilon_{\text{f}},\Phi)$ point is an independent root-finding problem for $(n,\mu_0)$. All runs converged to a unique, physical solution ($0<n<1$, finite $\kel{\Sigma}$ and $\kel{G}$ at every frequency), with no case of multiple attracting fixed points -- not even when the initial guess was changed for a few selected data points (not shown) -- or unphysical output anywhere in this work. This is consistent with the robust convergence already reported for the equilibrium KK-IPT scheme in Ref.~\cite{ka.ko.96}. However, we did find instances of non-convergence when the initial guess for $n$ was too close to either 0 or 1.}

\section{Results}\label{sec:results}

\subsection{Equilibrium benchmarks}

\begin{table}[b]
    \centering
    \begin{tabular}{@{\hspace{15pt}} c@{\hspace{15pt}} c@{\hspace{15pt}} c@{\hspace{15pt}} c@{\hspace{15pt}} c@{\hspace{15pt}}c@{\hspace{15pt}}}
        \toprule
        \text{Setup} & $U$ & $\epsilon_{\text{f}}$ & $T$ & $D$ & $\mu$ \\
        \midrule
        E1 & 5.5 & -3 & 0.05 & 10 & 0 \\
        E2 & 5.5 & 0 & 0.05 & 10 & 0 \\
        \bottomrule
    \end{tabular}
    \caption{The parameters for the two equilibrium setups discussed in the main text. The half-bandwidth of $\text{Im}\Delta^{\text{R}}(\omega)$ is denoted by $D$. All parameters are given in units of $\Gamma \equiv - \text{Im}\Delta^{\text{R}}(\omega = 0)$.}
    \label{tab:benchmarks_eq}
\end{table}

\begin{figure}[t]
\includegraphics[width=\linewidth]{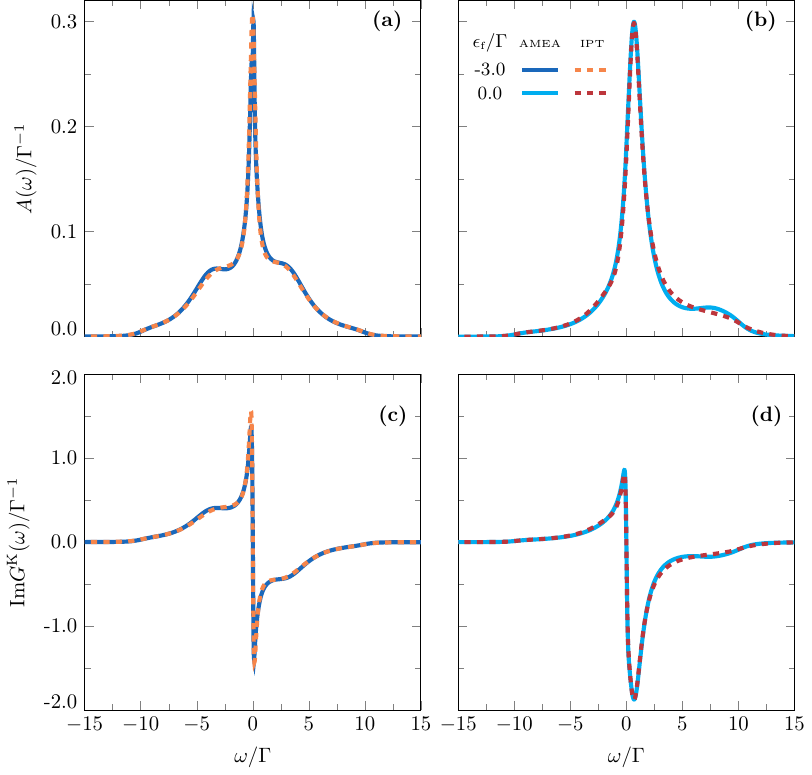}
\caption{Spectral functions for equilibrium setup (a) E1 ($\epsilon_{\text{\tiny f}} = -3\Gamma$) and (b) E2 ($\epsilon_{\text{\tiny f}} = 0$) obtained with AMEA and KK-IPT. (c) and (d) show the imaginary parts of the Keldysh components of the GF corresponding to (a) and (b), respectively. The occupations for setup E1 are $n_{\text{\tiny IPT}} = 0.5175$ and $n_{\text{\tiny AMEA}} = 0.5137$ while for E2 $n_{\text{\tiny IPT}} = 0.2584$ and $n_{\text{\tiny AMEA}} = 0.2626$. All relevant parameters can be found in Table~\ref{tab:benchmarks_eq}. For simplicity, the label KK-IPT has been shortened to IPT.}
\label{fig:fig1}
\end{figure}

\begin{figure}[b]
\includegraphics[width=\linewidth]{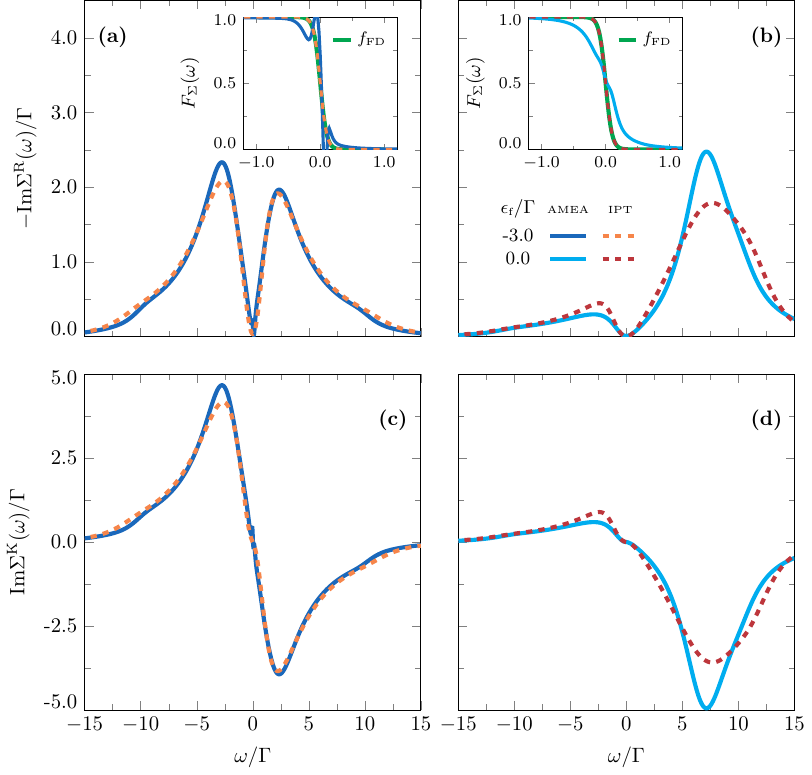}
\caption{Imaginary part of the retarded SE profiles for the equilibrium setups (a) E1 and (b) E2 obtained with AMEA and KK-IPT. The insets show the generalized distribution function of the SE, $F_{\Sigma}$, obtained with the two methods and how it compares to the Fermi-Dirac distribution function $f_{\text{\tiny FD}}$ with the same color code as in the main panels. (c) and (d) show the imaginary part of the Keldysh component of the SE corresponding to (a) and (b), respectively. All relevant parameters can be found in Table~\ref{tab:benchmarks_eq}. For simplicity, the label KK-IPT has been shortened to IPT.}
\label{fig:fig2}
\end{figure}

To start our analysis, we focus on equilibrium conditions and investigate the AIM in two different setups described by the parameters in Table~\ref{tab:benchmarks_eq}. Setup E1 corresponds to an impurity slightly more than half-filled while in E2 the impurity is roughly at quarter filling. As expected from~\cite{ka.ko.96}, the KK-IPT solver should provide accurate results for the present equilibrium case. And in fact, both setups show a nice agreement between results obtained by AMEA and KK-IPT, see Fig.~\ref{fig:fig1}. In particular, being close to half filling, the spectral function of setup E1 [Fig.~\ref{fig:fig1}(a)] still displays clearly visible lower and upper bands and the quasi-particle peak (QPP) at $\omega \approx 0$. On the other hand, the spectra in setup E2 [Fig.~\ref{fig:fig1}(b)] show a merge of the QPP with the lower band and a weak shoulder at $\omega \approx 8\Gamma$ corresponding to the upper band with both methods. The Keldysh component of the impurity GF corresponding to (a) and (b) is shown in Fig.~\ref{fig:fig1}(c) and (d), respectively.
Reproducing $G^{\text{K}}$ accurately is crucial to obtain the correct densities, as these are calculated from the lesser GF, see Eq.~\eqref{eq:occupation_num}. The exact values of the occupations for both methods are given in the caption of Fig.~\ref{fig:fig1}. To verify that the Ansatz~\eqref{eq:extended_ipt_se_ansatz} performs as intended, Fig.~\ref{fig:fig2}(a)-(d) shows the imaginary parts of the retarded and Keldysh components of the SEs corresponding to the spectra in the respective panels of Fig.~\ref{fig:fig1}. In light of the good agreement between the GFs produced by the two approaches, it is therefore unsurprising that the KK-IPT SE profiles qualitatively resemble those obtained with the AMEA impurity solver, particularly in the energy window around the chemical potential $\mu=0$. It should be noted, however, that the finite resolution of AMEA can introduce spurious features, since the auxiliary hybridization function $\underline{\Delta}_{\text{aux}}$ cannot reproduce the physical $\underline{\Delta}$ exactly~\footnote{For details on the AMEA impurity solver, we refer the reader to our previous work~\cite{do.nu.14,we.lo.23}.}. As a consequence, the generalized distribution function $F_{\Sigma} \equiv \left[ 1 - \mathrm{Im}\Sigma^{\mathrm{K}}/2\,\mathrm{Im}\Sigma^{\mathrm{R}}\right]/2$ of the SE obtained within AMEA may deviate from the Fermi-Dirac form even in equilibrium, as seen in the insets of Fig.~\ref{fig:fig2}(a) and (b). By contrast, because KK-IPT uses the physical $\underline{\Delta}$, it reproduces the Fermi-Dirac distribution exactly; see again the insets of Fig.~\ref{fig:fig2}(a) and (b). We also note that the generalized distributions for the GF and the SE, $F\equiv \left[ 1 - \mathrm{Im}G^{\mathrm{K}}/2\,\mathrm{Im}G^{\mathrm{R}}\right]/2$ and $F_{\Sigma}$, within AMEA may differ. This can be understood from the fact that the electronic SE is obtained as $\underline{\Sigma} = \underline{G}_0^{-1} - \underline{G}^{-1}$, where $\underline{G}_0$ is known exactly whereas $\underline{G}$ is approximate. This issue has already been noted in the literature, see, e.g., Ref.~\cite{bu.he.98}, and can lead to more pronounced artefacts in the SE distribution than in the one of the GF.

\subsection{Nonequilibrium steady state}\label{sec:neq_benchmarks}

In this section we consider electron transport across a correlated impurity between two non-interacting reservoirs under a finite bias voltage. In particular, we compare the differential conductance as obtained with AMEA and nonequilibrium KK-IPT. The current across the impurity is given by the Meir-Wingreen formula~\cite{me.wi.92}
\begin{equation}\label{eq:current}
J(\Phi, T) = \text{Re} \int \frac{\dd\omega}{2\pi} J(\omega; \Phi, T),
\end{equation}
where the integrand reads~\footnote{We point out that the retarded and Keldysh components of the impurity GF as well as the Keldysh component of the lead GF retain a non-trivial dependence on $\Phi$ and $T$. However, for our choice of leads, neither of these parameters enters the retarded (advanced) reservoir GF.}
\begin{equation}\label{eq:curr_integrand}
\begin{split}
J(\omega; \Phi, T) & = G^{\text{R}}(\omega; \Phi; T) \left( \Sigma^{\text{K}}_{\text{l}}(\omega; \Phi, T) - \Sigma^{\text{K}}_{\text{r}}(\omega; \Phi, T) \right) \\
& + G^{\text{K}}(\omega; \Phi, T) \left( \Sigma^{\text{R},\ast}_{\text{l}}(\omega) - \Sigma^{\text{R},\ast}_{\text{r}}(\omega) \right).
\end{split}
\end{equation}
In Eq.~\eqref{eq:curr_integrand} we have defined $\Sigma^{\text{R/K}}_{\lambda} \equiv t^{2}_{\lambda} L^{\text{R/K}}_{\lambda}$ and made the dependence on $\Phi$ and $T$ explicit. By exploiting the fact that the leads have the same density of states (see Eq.~\eqref{eq:leads_gf}) in~\eqref{eq:curr_integrand}, Eq.~\eqref{eq:current} can be simplified to
\begin{equation}
\begin{split}
J(\Phi, T) & = -\int \dd\omega \, A(\omega; \Phi, T) \text{Im}\Delta^{\text{R}}(\omega) \\
& \times \bigl( f_{\text{\tiny FD}}(\omega-\mu_{\text{l}}; T) - f_{\text{\tiny FD}}(\omega-\mu_{\text{r}}; T) \bigr),
\end{split}
\end{equation}
where we have used the definition of the spectral function
\begin{equation}\label{eq:spectral_function}
A(\omega) \equiv -\text{Im}G^{\text{R}}(\omega)/\pi.
\end{equation}
The leads' chemical potentials are shifted symmetrically in opposite directions by the application of the voltage bias, i.e. $\mu_{\text{l/r}} = \mu \pm \Phi/2$. The nonequilibrium conductance is then defined as
\begin{equation}\label{eq:conductance}
\mathcal{G}(\Phi, T) = \frac{\partial J(\Phi, T)}{\partial \Phi}.
\end{equation}
We compute numerically the nonequilibrium conductance~\eqref{eq:conductance} for selected interaction strengths $U$ and temperatures $T$ away from half filling. The temperatures used in our simulations are chosen as $T_{\text{K}}/4$ and $2T_{\text{K}}$, where $T_{\text{K}}$ denotes the Kondo temperature of the corresponding half-filled impurity at the same $U$ studied in Ref.~\cite{we.lo.23}. We note, however, that away from half filling the impurity occupancy changes, and thus the associated Kondo temperatures differ from those previously extracted in Ref.~\cite{we.lo.23} from NRG calculations. We nevertheless use those values of $T_{\text{K}}$ only as a reference scale to distinguish between low- and high-temperature regimes.

\begin{figure}[t]
\includegraphics[width=\linewidth]{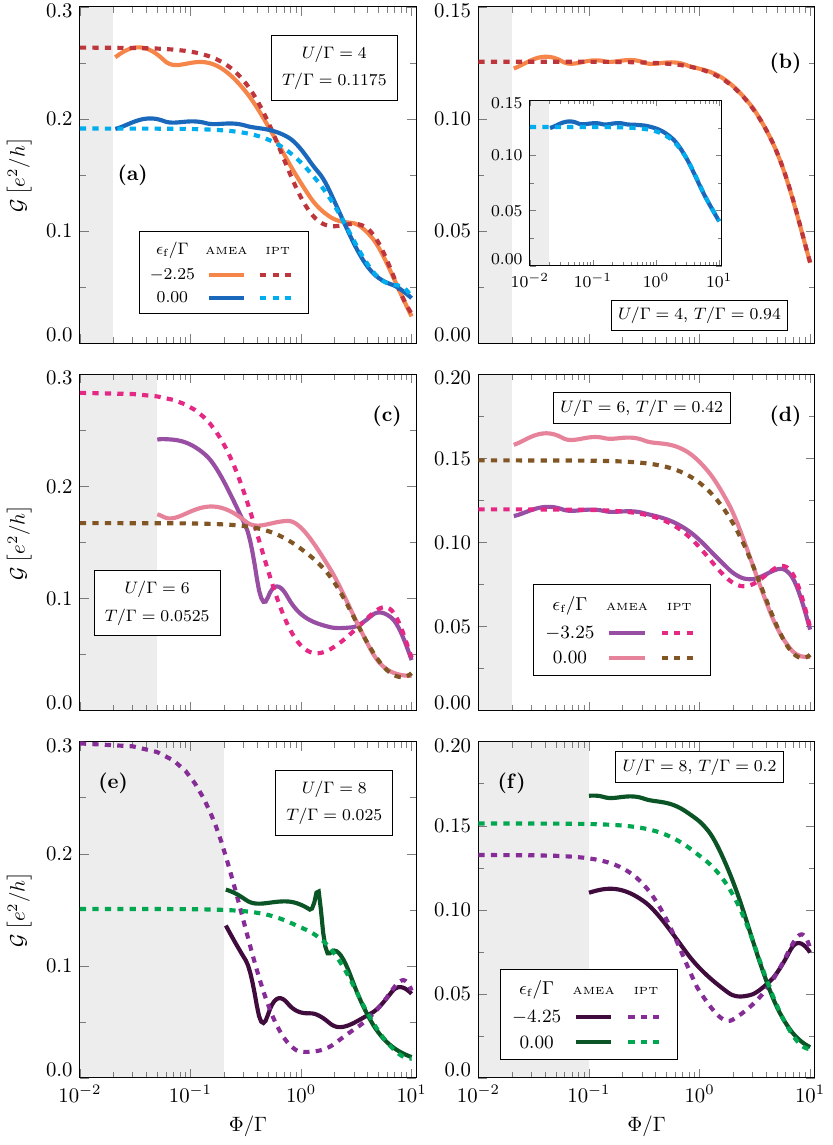}
\caption{Nonequilibrium differential conductance,  Eq.~\eqref{eq:conductance}, as function of the bias $\Phi$ for selected onsite energies $\epsilon_{\text{\tiny f}}$ obtained with AMEA and nonequilibrium KK-IPT. The upper, central, and lower rows correspond to $U=4\Gamma$, $6\Gamma$, and $8\Gamma$, respectively. [(a), (b)], [(c), (d)], and [(e), (f)] show $T/\Gamma=\left\{ 0.1175, 0.94\right\}$, $\left\{ 0.0525, 0.42\right\}$, and $\left\{0.025, 0.2\right\}$, respectively. The inset in (b) shows the case corresponding to $\epsilon_{\text{\tiny f}}=0$ alone while the shaded areas denote the voltage regions where $\mathcal{G}$ computed with AMEA is not reliable, see main text for the details. The voltage has been sampled according to $\Phi_{n} = \Phi_{0} 10^{n/150}$ with $n\in \left\{ 0, 1, 2, \dots, 150 \right\}$ and $\Phi_{0}=0.01 \Gamma$. For simplicity, the label KK-IPT has been shortened to IPT. (Here $D=10\Gamma$.)}
\label{fig:fig3}
\end{figure}

As already observed in Ref.~\cite{we.lo.23} at low temperatures ($T=T_{\text{K}}/4$), the conductance curves obtained with AMEA are reliable only down to moderately small voltages $\Phi$. For both fillings investigated in this work, the corresponding low-voltage region in which AMEA becomes unreliable grows with increasing Hubbard interaction $U$, reflecting the progressively smaller low-temperature scale; see Fig.~\ref{fig:fig3}(a), (c), and (e). This behavior can be traced back to the AMEA fitting procedure, the quality of which deteriorates at lower temperatures and in the presence of more intricate features of the hybridization function, such as those arising at finite but very small bias, where the Keldysh component develops a closely spaced double-step structure.

By contrast, nonequilibrium KK-IPT computes the SE from the physical hybridization function and therefore does not suffer from this fitting-related limitation. Consequently, it exhibits substantially fewer oscillations at small voltages and low temperatures, resulting in smoother conductance profiles for both fillings and every interaction strength considered here; see again Fig.~\ref{fig:fig3}(a), (c), and (e). Beyond these technical aspects, we emphasize the close agreement between the conductance curves obtained with AMEA and IPT, which remain in good qualitative and quantitative agreement throughout the voltage range where the AMEA results are reliable especially for small to intermediate values of the interaction $U$.

In Appendix~\ref{sec:occ_appendix}, we further validate nonequilibrium KK-IPT independently of AMEA by checking the exact first- and second-moment spectral sum rules, see Eqs.~\eqref{eq:first_moment} and~\eqref{eq:second_moment} therein; both are satisfied to better than $0.05\%$ across the entire parameter range studied, including the low-temperature, low-bias regime where a controlled AMEA benchmark is not available.

At larger temperatures ($T=2T_{\text{K}}$), the conductance curves obtained with AMEA remain reliable down to lower bias voltages $\Phi$ for small ($U=4\Gamma$) and intermediate ($U=6\Gamma$) interaction strengths than in the corresponding low-temperature regime; see Fig.~\ref{fig:fig3}(b) and (d). Only for the largest interaction considered, $U=8\Gamma$, do the AMEA results become more unstable, as seen in Fig.~\ref{fig:fig3}(f). Still, over the voltage range in which the AMEA results are reliable, nonequilibrium KK-IPT once again shows good qualitative and quantitative agreement with AMEA, in particular for small to intermediate values of $U$.

\begin{figure}[t]
\includegraphics[width=\linewidth]{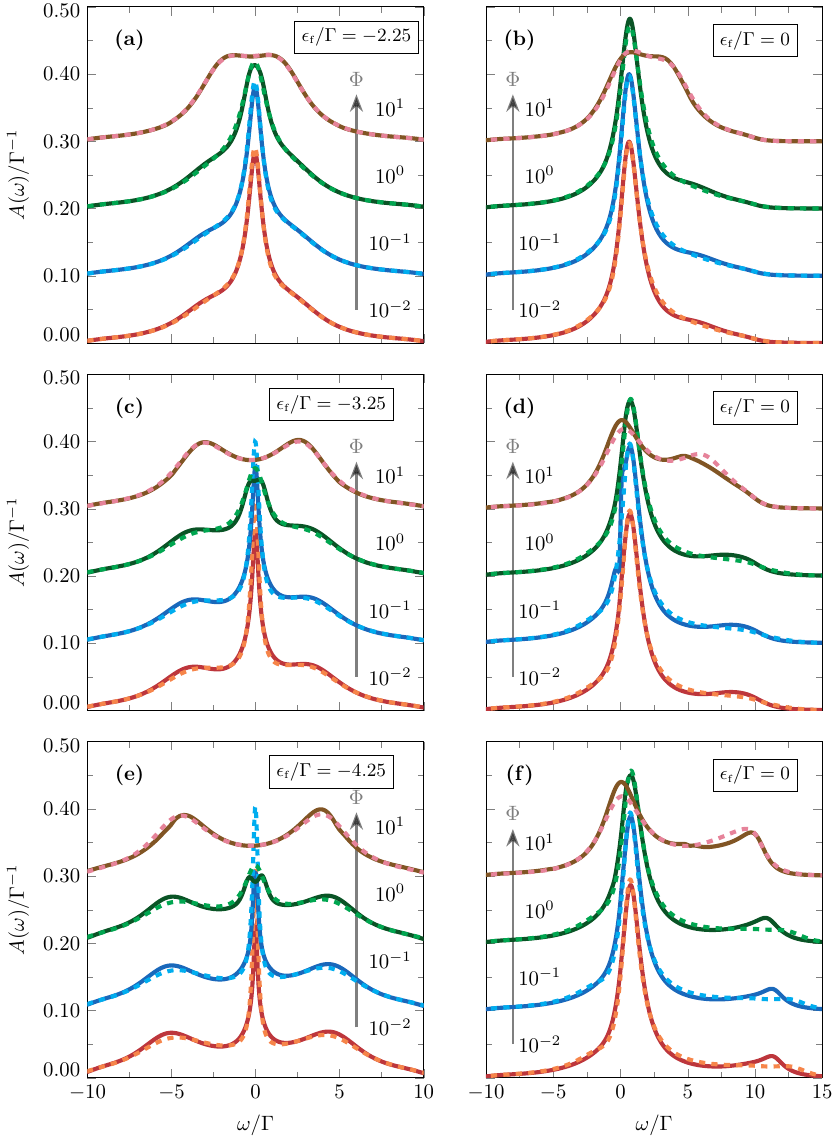}
\caption{Nonequilibrium spectral function \(A(\omega; \Phi, T)\) [Eq.~\eqref{eq:spectral_function}] for selected onsite energies \(\epsilon_{\text{\tiny f}}\) and bias voltages \(\Phi\), obtained with AMEA (solid lines) and nonequilibrium KK-IPT (dashed lines). Panels (a) and (b) show results at \(U=4\Gamma\) and \(T=0.1175\Gamma\). Panels (c) and (d) show results for \(U=6\Gamma\) and \(T=0.0525\Gamma\). Panels (e) and (f) show results for \(U=8\Gamma\) and \(T=0.025\Gamma\). Higher-temperature cases are not shown because thermal effects smear out the spectral features. Spectral functions corresponding to the increasing values of $\Phi$ have been shifted upward by a constant amount for better visualization. For simplicity, the label KK-IPT has been shortened to IPT. (Here \(D=10\Gamma\).)}
\label{fig:fig4}
\end{figure}

We conclude by discussing the nonequilibrium spectral features at low temperature, \(T=T_{\mathrm K}/4\), for selected values of the bias voltage which provide complementary information on the performance of nonequilibrium KK-IPT.

The spectra for \(T=2T_{\mathrm K}\) are not shown, because fine features are easily washed away by thermal broadening effects. For fillings slightly above \(1/2\) per spin, see panels (a), (c) and (e) in Fig.~\ref{fig:fig4}, the low-bias spectrum exhibits a three-peak structure, similar to the corresponding equilibrium one, consisting of lower and upper bands separated by approximately \(U\) and a Kondo resonance at \(\omega \approx 0\). For \(U=6\Gamma\) and \(8\Gamma\), this structure is clearly resolved, whereas for \(U=4\Gamma\) the bands appear only as {\em shoulders} of the central Kondo peak; see Fig.~\ref{fig:fig4}(a), (c), and (e).

As the interaction strength \(U\) is increased, the Kondo resonance at \(\omega \approx 0\) exhibits bias-induced broadening and central suppression consistent with the onset of peak splitting at progressively lower bias voltages \(\Phi\); see again Fig.~\ref{fig:fig4}(a), (c), and (e). This behavior reflects the fact that larger values of \(U\) require lower temperatures to resolve a sufficiently sharp Kondo peak, which is in turn more susceptible to suppression by an applied bias voltage; see, e.g., Ref.~\cite{we.lo.23}. Overall, the spectra obtained with nonequilibrium KK-IPT are in very good agreement with those obtained with AMEA for all interaction strengths \(U\) considered in this work, at this filling. However, for \(U=6\Gamma\) and \(8\Gamma\), the Kondo peak obtained with IPT is noticeably more pronounced than in AMEA at \(\Phi=0.1\Gamma\); see Fig.~\ref{fig:fig4}(c) and (e). This discrepancy may be related to the aforementioned limitations of AMEA at low bias, since the agreement improves again as the voltage is increased.

For fillings well away from \(1/2\), see the panels on the right-hand side of Fig.~\ref{fig:fig4}, the agreement between nonequilibrium KK-IPT and AMEA at low bias remains good for small and intermediate values of \(U\); see Fig.~\ref{fig:fig4}(b) and (d). In this regime, nonequilibrium KK-IPT satisfactorily reproduces both the low-energy peak, which results from the merging of the Kondo resonance with the lower band, and the upper band, which appears as a shoulder at an energy roughly \(U\) above the first peak. For larger values of \(U\), however, nonequilibrium KK-IPT underestimates the amplitude of the upper band, tending to smear it out and to transfer part of the spectral weight to higher energies; see Fig.~\ref{fig:fig4}(f). As the bias \(\Phi\) is further increased, the low-energy peak develops a pronounced transfer of spectral weight toward higher energies, which is well captured by nonequilibrium KK-IPT. This effect becomes more marked with increasing \(U\). As a result, the upper band appears to merge progressively with the high-energy tail of the low-energy peak; see again Fig.~\ref{fig:fig4}(b), (d), and (f). 

The nonequilibrium behavior of the particle number and double occupancy, discussed in detail in Appendix~\ref{sec:occ_appendix}, follows a similar pattern to the spectral functions and conductance above. The particle number agrees well between nonequilibrium KK-IPT and AMEA across the parameter range studied, with the largest deviations occurring away from half filling at strong interactions. 
The double occupancy, however, is a more demanding test: near half filling and within the bias window $\Phi/\Gamma\in[1,10]$ where AMEA is reliable, the two methods agree to within  $\sim 10^{-3}$, but the discrepancy grows to $\sim 10^{-2}$ away from half filling, particularly for $U\gtrsim6\Gamma$ (Table~\ref{tab:nd_error}). This regime dependence, together with the exact spectral sum rules discussed above, is used in Appendix~\ref{sec:occ_appendix} to delimit quantitatively where nonequilibrium KK-IPT can be trusted beyond the single-particle level.

\section{Conclusions}\label{sec:conclusions}

In this work, we extend the Ansatz for the electronic self-energy (SE) proposed by Kajeuter and Kotliar within the iterated perturbation theory (KK-IPT), see Ref.~\cite{ka.ko.96}, to build an approximate nonequilibrium single-impurity solver that works beyond half filling. We then apply the resulting nonequilibrium KK-IPT to the Anderson impurity problem under the influence of a bias voltage.

As a first validation step, we benchmark the method in equilibrium for two representative setups, corresponding to an impurity slightly above half filling and to roughly quarter filling. In both cases, the  KK-IPT reproduces the results obtained with the auxiliary master equation approach (AMEA) accurately at the level of both the spectral function and the Keldysh component of the Green's function, and therefore also yields the correct impurity occupations. In addition, the retarded and Keldysh components of the KK-IPT SE closely follow the corresponding AMEA results, especially in the energy window around the chemical potential. While the finite resolution of AMEA may induce spurious effects and deviations from the exact Fermi-Dirac expression in the generalized distribution function of the SE, KK-IPT, which is formulated directly in terms of the physical hybridization function, reproduces the equilibrium distribution exactly.

Building on this equilibrium benchmark, we then apply the nonequilibrium KK-IPT impurity solver to nonequilibrium steady-state transport across a correlated impurity away from half filling and analyze the differential conductance for different interaction strengths, temperatures, and fillings. Over the voltage range in which the AMEA results are reliable, nonequilibrium KK-IPT shows good qualitative and quantitative agreement with AMEA, reproducing the main transport features across the parameter regime studied here. This agreement is further supported by the nonequilibrium spectral functions at low temperature, which provide complementary information on the performance of the method. Near half filling and for low bias voltages nonequilibrium KK-IPT reproduces the characteristic three-peak structure consisting of lower and upper bands and a low-energy Kondo resonance, as well as its evolution under increasing bias. Moving away from near half filling, nonequilibrium KK-IPT also captures the bias-induced redistribution of spectral weight toward higher energies and the corresponding progressive loss of separation between the low-energy peak and the upper band, although for larger interaction strengths it tends to underestimate the weight of the latter at low biases. This overall agreement is particularly notable given the approximate character of nonequilibrium KK-IPT compared to the numerical effort required by AMEA. At the same time, the comparison highlights an important practical difference between the two approaches. In the low-bias, low-temperature regime, the relevant low-energy features become increasingly difficult to resolve within AMEA, whereas nonequilibrium KK-IPT remains numerically stable and the conductance then displays substantially smoother behavior. Although a controlled benchmark is not available in that regime, this indicates that this nonequilibrium KK-IPT may provide a fast, albeit qualitative, alternative for nonequilibrium steady-state calculations away from half filling, in particular at low temperatures and biases, in the paramagnetic regime.

We stress that the above results should be read with a clear distinction between the regime in which they can be rigorously benchmarked against AMEA and the regime in which they cannot. Where AMEA itself constitutes a controlled reference -- moderate-to-large bias and, depending on $U$, moderate-to-low temperature~\footnote{The precise range of validity depends on both $U$ and $T$; see the shaded regions in Fig.~\ref{fig:fig3}. At $U=4\Gamma$, AMEA is reliable down to $\Phi/\Gamma\approx0.02$, essentially independently of temperature (cf.\ also Ref.~\cite{we.lo.23}, Fig.~6 therein, which suggests that the conductance $\mathcal{G}$ -- itself obtained by numerical derivative, and thus error-amplifying -- is accurate for $T/T_K\lesssim10^{-2}$ at intermediate $U/\Gamma\sim4$). At $U=6\Gamma$, the reliable range extends down to $\Phi/\Gamma\approx0.05$ at low temperature ($T=T_K/4$) and $\Phi/\Gamma\approx0.02$ at high temperature ($T=2T_K$). At $U=8\Gamma$, it extends down to $\Phi/\Gamma\approx0.2$ at low temperature and $\Phi/\Gamma\approx0.1$ at high temperature. In short, the bias threshold below which AMEA becomes unreliable grows sharply with $U$, while temperature plays a secondary role that only becomes significant once $U\gtrsim6\Gamma$.} -- nonequilibrium KK-IPT is not only qualitatively but quantitatively accurate, reproducing even the double occupancy to within $\sim 10^{-3}$ near half filling (Appendix~\ref{sec:occ_appendix}). This is a nontrivial finding: the IPT-$n_0$ self-consistency loop only constrains the single-particle occupation $n$ [Eq.~\eqref{eq:occupation_num}]; nothing in its construction guarantees that $n_{\text{d}}$, a genuinely two-particle observable, is reproduced with comparable accuracy.

By contrast, at small bias and low temperature, where the AMEA fitting procedure itself becomes unreliable, we cannot draw an equally firm conclusion. Here KK-IPT and AMEA particle numbers still agree well; however, this agreement is not conclusive, since in the absence of a controlled benchmark it cannot exclude that both methods share a common bias, and single-particle occupations are in any case less sensitive to the approximations involved than higher-order observables. This benchmark can therefore quantitatively validate nonequilibrium KK-IPT only within the regime accessible to AMEA; beyond it, and especially away from half filling at strong coupling, the comparison is only qualitative or semi-quantitative.

A separate and complementary piece of evidence is that the exact first- and second-moment spectral sum rules, Eqs.~\eqref{eq:first_moment} and \eqref{eq:second_moment}, are satisfied to better than $0.05\%$ throughout the entire parameter range studied, including the low-temperature, low-bias regime where no AMEA benchmark exists, see Appendix~\ref{sec:occ_appendix}. This should, however, be read as a necessary rather than a sufficient condition for the reliability of nonequilibrium KK-IPT. Both sum rules are asymptotic, large-$\omega$ statements about the retarded self-energy alone: the first moment follows from the exact Hartree limit $\Sigma^{\text{R}}(\omega\to\infty)\to Un$, already enforced by construction, and the second from the atomic-limit coefficient $U^2n(1-n)$, which likewise depends only on $n$ and not on the quality of the diagrammatic resummation at finite frequency. The occupation $n$ entering both coefficients is of course obtained from the full nonequilibrium solution and therefore is not independent of the Keldysh sector; but the sum rules constrain it only as a single, aggregate number consistent with the asymptotic tail of $\Sigma^{\text{R}}(\omega)$, not as the frequency-resolved $\Sigma^{\text{K}}(\omega)$ and $G^{\text{K}}(\omega)$ that set the mid- and low-frequency lineshape entering observables such as the double occupancy. Consequently, the excellent agreement of the moments with their exact values -- essentially guaranteed by the structure of the Ansatz -- does not by itself imply comparable accuracy for $n_{\text{d}}$, which requires the full frequency dependence of \emph{both} the retarded and the lesser/Keldysh sectors to be captured correctly, and which we have shown deteriorates markedly away from half filling. We regard this combination -- exact sum rules holding throughout, together with a controlled and quantitatively delimited regime of validity for higher-order observables -- as giving a realistic assessment of nonequilibrium KK-IPT: mathematically consistent and numerically stable everywhere, but quantitatively trustworthy for observables beyond the single-particle level only near half filling and within the bias/temperature regime accessible to AMEA. We nonetheless find this encouraging, as it suggests that nonequilibrium KK-IPT may prove useful within genuine nonequilibrium DMFT calculations, where numerically exact impurity solvers are often out of reach or numerically very expensive, provided its predictions for higher-order observables are interpreted with the caveats established here.

Overall, our results identify nonequilibrium KK-IPT as a computationally inexpensive impurity solver for the equilibrium and nonequilibrium steady-state problems studied here at arbitrary filling, and motivate further applications to other observables and correlated systems and, especially within dynamical mean-field theory. We stress, however, that the present nonequilibrium construction may not be unique. In equilibrium, for instance, several variants of IPT have been considered, including the IPT-$n_0$ approximation and schemes enforcing the Friedel sum rule. Since no direct nonequilibrium counterpart of the Friedel sum rule is available, we adopted here the simplest possible extension, namely a nonequilibrium generalization of IPT-$n_0$. The results obtained in this work show that this choice is viable and performs satisfactorily for the cases considered. Nevertheless, a broader assessment is required to determine its range of applicability, in particular in more challenging situations such as structured or gapped hybridization functions or non-equivalent spin species, and to clarify whether alternative nonequilibrium KK-IPT constructions may offer additional advantages. Work in this direction is currently underway.

\appendix

\section{Derivation of the Keldysh component of the electronic self-energy}
\label{sec:keld_se_derivation}

In this appendix we derive the Keldysh component of the electronic self-energy in Eq.~\eqref{eq:extended_ipt_se_ansatz}. For objects with the Keldysh structure defined in Eq.~\eqref{eq:keld_struct}, the ordering of factors matters. One may therefore wonder whether the two orderings
\begin{equation}
\label{eq:candidates_keld}
\begin{split}
\underline{\Sigma}^{(1)} & = \underline{\Sigma}_{\text{\tiny A}} \left( 1 - \underline{\Sigma}_{\text{\tiny B}}\right)^{-1}, \\
\underline{\Sigma}^{(2)} & = \left( 1 - \underline{\Sigma}_{\text{\tiny B}}\right)^{-1} \underline{\Sigma}_{\text{\tiny A}},
\end{split}
\end{equation}
lead to different Keldysh components. Here and in the following, we suppress the frequency argument for notational simplicity.

For two generic Keldysh objects \(\underline{X}\) and \(\underline{Y}\), the multiplication rule reads
\begin{equation}
\label{eq:kel_mult}
\left( \underline{X}\underline{Y} \right)^{\text{K}} = X^{\text{R}} Y^{\text{K}} + X^{\text{K}} Y^{\text{A}}.
\end{equation}
Hence, the order of the factors in Eq.~\eqref{eq:candidates_keld} is in principle relevant. We also recall that the Keldysh component of the inverse is
\begin{equation}
\label{eq:inv_kel}
\left( X^{-1} \right)^{\text{K}} = - \left( X^{-1} \right)^{\text{R}} X^{\text{K}} \left( X^{-1} \right)^{\text{A}},
\end{equation}
with \(\left( X^{-1} \right)^{\text{R/A}} = \left( X^{\text{R/A}} \right)^{-1}\). Extracting the Keldysh components of the two expressions in Eq.~\eqref{eq:candidates_keld}, and using Eqs.~\eqref{eq:kel_mult} and \eqref{eq:inv_kel}, we obtain
\begin{equation}
\label{eq:candidates_keld_comps}
\begin{split}
{\Sigma^{(1)}}^{\text{K}} & =
\left[
\Sigma^{\text{K}}_{\text{\tiny A}}
+
\Sigma^{\text{R}}_{\text{\tiny A}}
\left( 1 - \Sigma^{\text{R}}_{\text{\tiny B}} \right)^{-1}
\Sigma^{\text{K}}_{\text{\tiny B}}
\right]
\left( 1 - \Sigma^{\text{A}}_{\text{\tiny B}} \right)^{-1}, \\
{\Sigma^{(2)}}^{\text{K}} & =
\left( 1 - \Sigma^{\text{R}}_{\text{\tiny B}}\right)^{-1}
\left[
\Sigma^{\text{K}}_{\text{\tiny A}}
+
\Sigma^{\text{K}}_{\text{\tiny B}}
\left( 1 - \Sigma^{\text{A}}_{\text{\tiny B}} \right)^{-1}
\Sigma^{\text{A}}_{\text{\tiny A}}
\right].
\end{split}
\end{equation}
The second line of Eq.~\eqref{eq:candidates_keld_comps} is the expression used in Eq.~\eqref{eq:extended_ipt_se_ansatz}.

At first sight, the two expressions in Eq.~\eqref{eq:candidates_keld_comps} need not coincide. We now show that since $\underline{\Sigma}_{\text{\tiny A}}$ and $\underline{\Sigma}_{\text{\tiny B}}$ are proportional, see Eq.~\eqref{eq:shorthand_notation_coeffs}, they are in fact identical. Using the definitions in Eq.~\eqref{eq:shorthand_notation_coeffs}, the first line of Eq.~\eqref{eq:candidates_keld_comps} can be rewritten as
\begin{equation}
\label{eq:k1_intermediate}
{\Sigma^{(1)}}^{\text{K}} =
A
\left[
1 + B \tilde{\Sigma}^{\text{R}} \left( 1 - B \tilde{\Sigma}^{\text{R}} \right)^{-1}
\right]
\tilde{\Sigma}^{\text{K}}
\left( 1 - B \tilde{\Sigma}^{\text{A}} \right)^{-1}.
\end{equation}
Using the identity
\[
1 + B \tilde{\Sigma}^{\text{R}} \left( 1 - B \tilde{\Sigma}^{\text{R}} \right)^{-1}
=
\left( 1 - B \tilde{\Sigma}^{\text{R}} \right)^{-1},
\]
we obtain
\begin{equation}
\label{eq:k1_reduced_general}
{\Sigma^{(1)}}^{\text{K}} =
A
\left( 1 - B \tilde{\Sigma}^{\text{R}} \right)^{-1}
\tilde{\Sigma}^{\text{K}}
\left( 1 - B \tilde{\Sigma}^{\text{A}} \right)^{-1}.
\end{equation}

Similarly, the second line of Eq.~\eqref{eq:candidates_keld_comps} becomes
\begin{equation}
\label{eq:k2_intermediate}
{\Sigma^{(2)}}^{\text{K}} =
A
\left( 1 - B \tilde{\Sigma}^{\text{R}} \right)^{-1}
\tilde{\Sigma}^{\text{K}}
\left[
1 + \left( 1 - B \tilde{\Sigma}^{\text{A}} \right)^{-1} B \tilde{\Sigma}^{\text{A}}
\right].
\end{equation}
Employing the identity
\[
1 + \left( 1 - B \tilde{\Sigma}^{\text{A}} \right)^{-1} B \tilde{\Sigma}^{\text{A}}
=
\left( 1 - B \tilde{\Sigma}^{\text{A}} \right)^{-1},
\]
we find
\begin{equation}
\label{eq:k2_reduced_general}
{\Sigma^{(2)}}^{\text{K}} =
A
\left( 1 - B \tilde{\Sigma}^{\text{R}} \right)^{-1}
\tilde{\Sigma}^{\text{K}}
\left( 1 - B \tilde{\Sigma}^{\text{A}} \right)^{-1}.
\end{equation}
Equations~\eqref{eq:k1_reduced_general} and \eqref{eq:k2_reduced_general} are therefore identical, showing that the Keldysh component is uniquely defined in the case in which $\underline{\Sigma}_{\text{\tiny A}}$ and $\underline{\Sigma}_{\text{\tiny B}}$ are proportional.

For the scalar complex-valued functions considered in this work, Eq.~\eqref{eq:k1_reduced_general} reduces to
\begin{equation}
\label{eq:k_scalar}
{\Sigma}^{\text{K}} =
\frac{A \tilde{\Sigma}^{\text{K}}}{\left| 1 - B \tilde{\Sigma}^{\text{R}} \right|^{2}}.
\end{equation}
Since the Keldysh component is anti-Hermitian, \(\tilde{\Sigma}^{\text{K}}\) is purely imaginary in the scalar case, and therefore so is \({\Sigma}^{\text{K}}\).

The form in Eq.~\eqref{eq:k1_reduced_general} also remains valid for matrix-valued quantities, for instance in orbital space. In that case, however, anti-Hermiticity implies
\[
\left( {\bm{\Sigma}}^{\text{K}} \right)^{\dagger} = - {\bm{\Sigma}}^{\text{K}},
\]
where the bold letters denote matrices, rather than elementwise pure imaginary character.

\section{Occupations}\label{sec:occ_appendix}

\subsection{Single-particle occupations}

In order to complete our analysis of the nonequilibrium setups discussed in Sec.~\ref{sec:neq_benchmarks}, we present here the impurity occupation and double occupancy as functions of the applied bias voltage for the interaction strengths and temperatures considered in Fig.~\ref{fig:fig3}.
\begin{figure}[t]
\includegraphics[width=\linewidth]{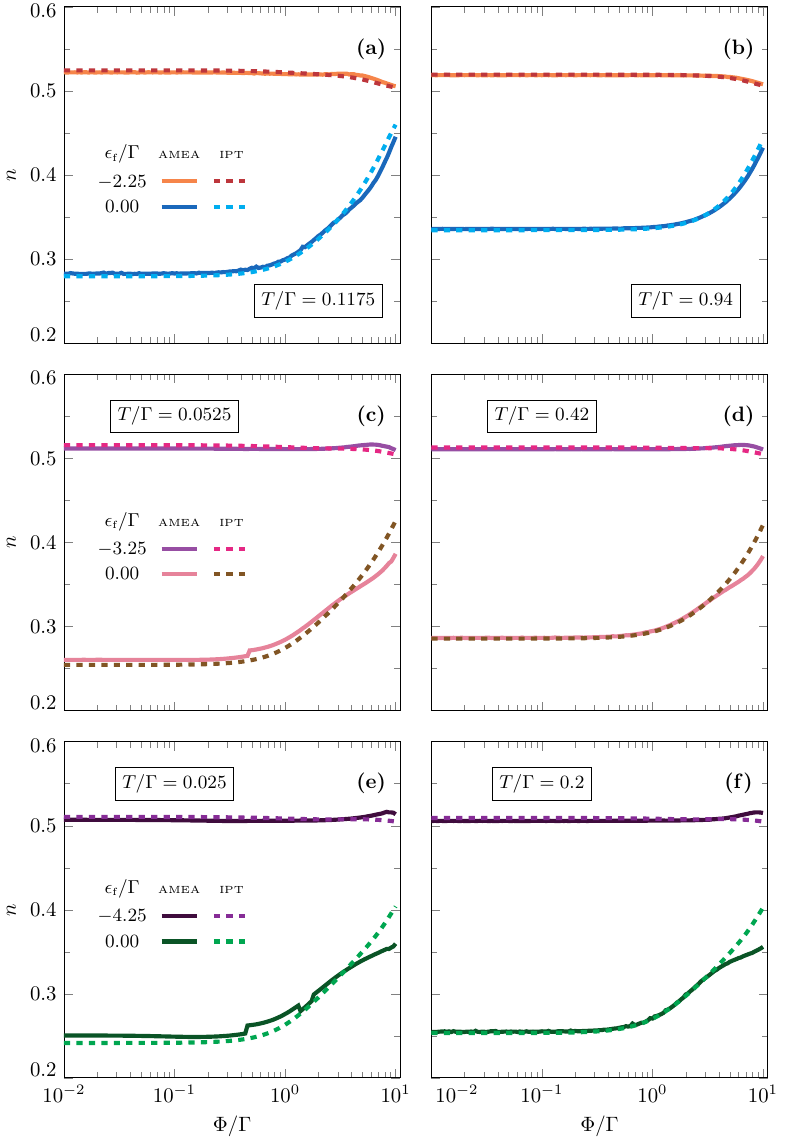}
\caption{Particle number, see Eq.~\eqref{eq:occupation_num}, as function of the bias $\Phi$ for the setups discussed in Fig.~\ref{fig:fig3}. The upper row [panels (a) and (b)] shows the results for $U=4\Gamma$, the central row [(c) and (d)] for $U=6\Gamma$ and the lower one [(e) and (f)] for $U=8\Gamma$. Temperatures and onsite energies are reported on the plots. For simplicity, the label KK-IPT has been shortened to IPT.}
\label{fig:fig5}
\end{figure}
Figure~\ref{fig:fig5} shows the impurity occupation for $U=4\Gamma$ in panels (a) and (b), for $U=6\Gamma$ in panels (c) and (d), and for $U=8\Gamma$ in panels (e) and (f). As in Fig.~\ref{fig:fig3}, the left panels correspond to low temperature, while the right panels show the high-temperature results.

For the weakest interaction, $U=4\Gamma$, shown in Fig.~\ref{fig:fig5}(a) and (b), the agreement between nonequilibrium KK-IPT and AMEA is very good for all fillings, and is particularly good at high temperature. For intermediate and strong interactions, $U=6\Gamma$ and $U=8\Gamma$, the agreement tends to worsen, especially at large bias voltages. In all cases, the largest deviations between the two methods occur for the parameter set furthest from half filling, most notably at intermediate and strong coupling, see Fig.~\ref{fig:fig5}(c)-(f). A common trend in all setups is that, at large bias voltage $\Phi$, the occupation of the case furthest from half filling increases, whereas the one closer to half filling remains approximately constant.
\begin{figure}[t]
\includegraphics[width=\linewidth]{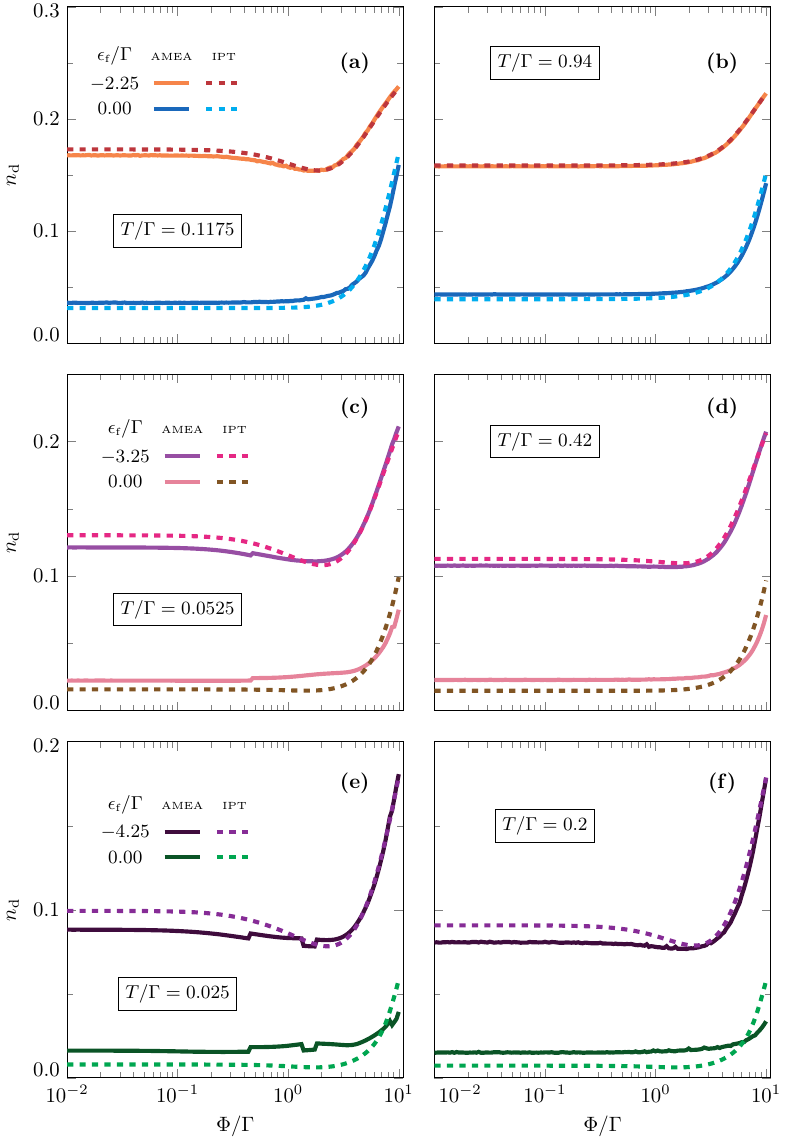}
\caption{Double occupancy, see Eq.~\eqref{eq:double_occupancy}, as function of the bias $\Phi$ for the setups discussed in Fig.~\ref{fig:fig3}. The upper row [panels (a) and (b)] shows the results for $U=4\Gamma$, the central row [(c) and (d)] for $U=6\Gamma$ and the lower one [(e) and (f)] for $U=8\Gamma$. Temperatures and onsite energies are reported on the plots. For simplicity, the label KK-IPT has been shortened to IPT.}
\label{fig:fig6}
\end{figure}

\subsection{Double occupancy}

We estimate the double occupancy via the exact nonequilibrium Galitskii-Migdal expression~\cite{st.va.13}
\begin{equation}\label{eq:double_occupancy}
n_{\text{d}} = \frac{-\ii}{2\pi U} \int \dd \omega \left[ \Sigma^{\text{R}}(\omega) G^{<}(\omega) + \Sigma^{<}(\omega) G^{\text{A}}(\omega) \right],
\end{equation}
with the lesser components obtained from Eq.~\eqref{eq:lssgtr_from_RK} and $G^{\text{A}} = \left( G^{\text{R}} \right)^{\ast}$.

The double occupancy exhibits a similar overall trend to the particle number, see Fig.~\ref{fig:fig6}. Unlike for the particle number, however, the two fillings considered for each $(U,T)$ show markedly different behavior: near half filling, nonequilibrium KK-IPT and AMEA agree closely in the region $\Phi/\Gamma \in [1,10]$, regardless of the temperature considered, where AMEA is most reliable, see the discussion of Fig.~\ref{fig:fig3}.

By contrast, away from half filling the agreement worsens considerably and deteriorates further as $U$ is increased, already at $\Phi/\Gamma \in [1,10]$ rather than at smaller biases. Table~\ref{tab:nd_error} quantifies this trend via the mean absolute deviation $|n_{\text{d}}^{\text{\tiny IPT}} - n_{\text{d}}^{\text{\tiny AMEA}}|$ restricted to $\Phi/\Gamma \in [1,10]$, i.e. where AMEA is reliable. We report the absolute rather than the relative deviation because, away from half filling, $n_{\text{d}}^{\text{\tiny AMEA}}$ itself is small, which would otherwise inflate the error even when the absolute discrepancy is comparable to that seen near half filling.

The mean absolute deviation remains small, below $\sim0.004$, near half filling for every $U$ and $T$ studied. Away from half filling it is systematically larger, by a factor of $2$--$8$ at fixed $(U,T)$: it is comparatively modest at $U=4\Gamma$ ($\sim0.004$--$0.007$), then increases and plateaus around $0.009$--$0.010$ for $U=6\Gamma$ and $8\Gamma$, with little further growth or temperature dependence beyond that point.

\begin{figure}[t]
\includegraphics[width=\linewidth]{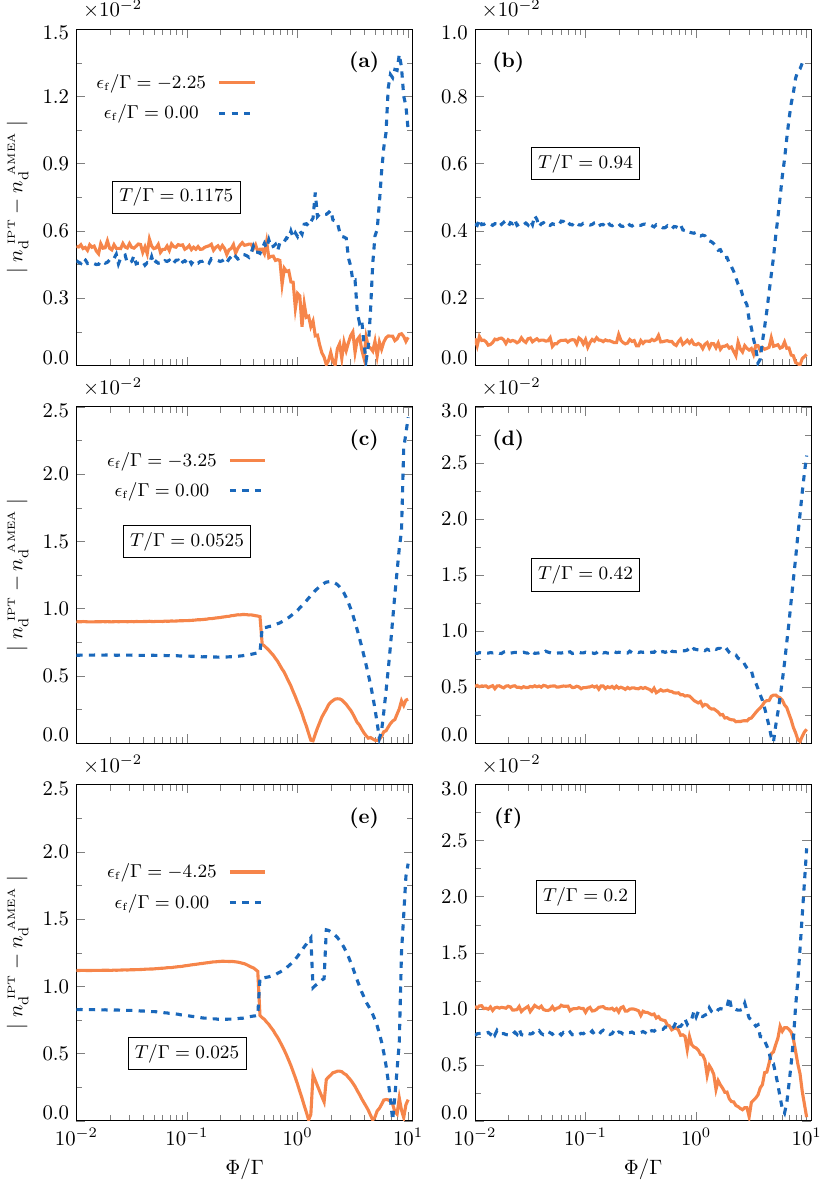}
\caption{Absolute deviation of the double occupancy, $|n_{\text{d}}^{\text{\tiny IPT}} - n_{\text{d}}^{\text{\tiny AMEA}}|$, as a function of bias $\Phi$ for the setups discussed in Fig.~\ref{fig:fig3}. The upper row shows the results for $U=4\Gamma$, the central row for $U=6\Gamma$ and the lower one for $U=8\Gamma$. Temperatures and onsite energies are reported on the plots.}
\label{fig:error_nd}
\end{figure}

Figure~\ref{fig:error_nd} shows the corresponding absolute deviation explicitly as a function of bias. These results indicate that the quantitative accuracy of nonequilibrium KK-IPT for higher-order observables such as the double occupancy is largely restricted to the vicinity of half filling and for $\Phi/\Gamma \in [1,10]$, and should be regarded as qualitative (i) in the region where AMEA is less reliable (independent of the filling) and (ii) away from half-filled settings, particularly for $U \gtrsim 6\Gamma$.

\begin{table}[t]
\centering
\footnotesize
\begin{tabular}{@{\hspace{5pt}} c@{\hspace{5pt}} c@{\hspace{5pt}} c@{\hspace{5pt}} l@{\hspace{5pt}} c@{\hspace{5pt}}}
\toprule
$U/\Gamma$ & $T/\Gamma$ & $\epsilon_{\text{f}}/\Gamma$ & filling & mean abs.\ deviation \\
\midrule
4 & 0.1175 & $-2.25$ & near half filling & 0.00113 \\
4 & 0.1175 & $0$     & away from half filling & 0.00692 \\
4 & 0.94   & $-2.25$ & near half filling & 0.00049 \\
4 & 0.94   & $0$     & away from half filling & 0.00392 \\
6 & 0.0525 & $-3.25$ & near half filling & 0.00179 \\
6 & 0.0525 & $0$     & away from half filling & 0.00986 \\
6 & 0.42   & $-3.25$ & near half filling & 0.00266 \\
6 & 0.42   & $0$     & away from half filling & 0.00850 \\
8 & 0.025  & $-4.25$ & near half filling & 0.00190 \\
8 & 0.025  & $0$     & away from half filling & 0.01028 \\
8 & 0.2    & $-4.25$ & near half filling & 0.00409 \\
8 & 0.2    & $0$     & away from half filling & 0.00895 \\
\bottomrule
\end{tabular}
\caption{Mean absolute deviation of the double occupancy, $|n_{\text{d}}^{\text{\tiny IPT}} - n_{\text{d}}^{\text{\tiny AMEA}}|$, Eq.~\eqref{eq:double_occupancy}, restricted to the bias window $\Phi/\Gamma \in [1,10]$ where AMEA is reliable; see also Fig.~\ref{fig:fig3} and accompanying discussion. "Near half filling'' and "away from half filling'' refer to the two onsite energies $\epsilon_{\text{f}}$ studied for each $(U,T)$ in Figs.~\ref{fig:fig3}--\ref{fig:fig6}.}
\label{tab:nd_error}
\end{table}

\subsection{Spectral moment sum rules}\label{sec:sum_rules}

The retarded Green's function $G^{\text{R}}(\omega)$ admits an exact expansion in inverse powers of $\omega$ at large frequency, with coefficients (moments) given by equal-time expectation values of nested commutators of $\hat{f}_{\sigma}$ with the total Hamiltonian $\hat{H}$ in Eq.~\eqref{eq:tot_hamiltonian}. This moment-expansion technique originates in Zubarev's equation-of-motion method for double-time Green's functions~\cite{zuba.60}, was first applied by Harris and Lange to derive spectral moment sum rules for the Hubbard model~\cite{ha.la.67}, and was later systematized by Nolting~\cite{nolt.72}.

The corresponding equilibrium self-energy moments for the single-band Anderson impurity model are likewise well established~\cite{de.he.97,po.we.97,bu.he.98,wa.da.11}.

Crucially, the derivation relies only on the equal-time canonical anticommutation relations and the Heisenberg equation of motion, and is therefore a state-independent operator identity that holds in any steady state, equilibrium or not -- a property already exploited for driven lattice models in Ref.~\cite{tu.fr.08}. To our knowledge, the extension given below to the genuine two-lead nonequilibrium steady state of a quantum impurity has not been made explicit before. For the model considered here, the first two moments read
\begin{equation}\label{eq:first_moment}
M_{1} = \epsilon_{\text{f}} + Un = \int \dd \omega \, \omega \, A(\omega),
\end{equation}
\begin{equation}\label{eq:second_moment}
M_{2} = \left( M_{1} \right)^{2} + U^{2}n(1-n) + D_{1} = \int \dd \omega \, \omega^{2} \, A(\omega),
\end{equation}
The term $D_{1} \equiv - \frac{1}{\pi} \int \dd\omega \, \text{Im}\Delta^{\text{R}}(\omega)$ is the total spectral weight ({\em zeroth} moment) of the hybridization function, i.e.\ the leading coefficient in its large-$\omega$ expansion $\Delta^{\text{R}}(\omega) = D_1/\omega + O(1/\omega^2)$. The first moment, Eq.~\eqref{eq:first_moment}, follows directly from the exact Hartree limit $\Sigma^{\text{R}}(\omega \to \infty) \to Un$, while the $U^{2}n(1-n)$ term in the second moment is the well-known atomic-limit second self-energy moment~\cite{po.we.97,bu.he.98}.

As we are considering $n_{\uparrow} = n_{\downarrow} = n$, we dropped the spin quantum number.

\begin{figure}[t]
\includegraphics[width=\linewidth]{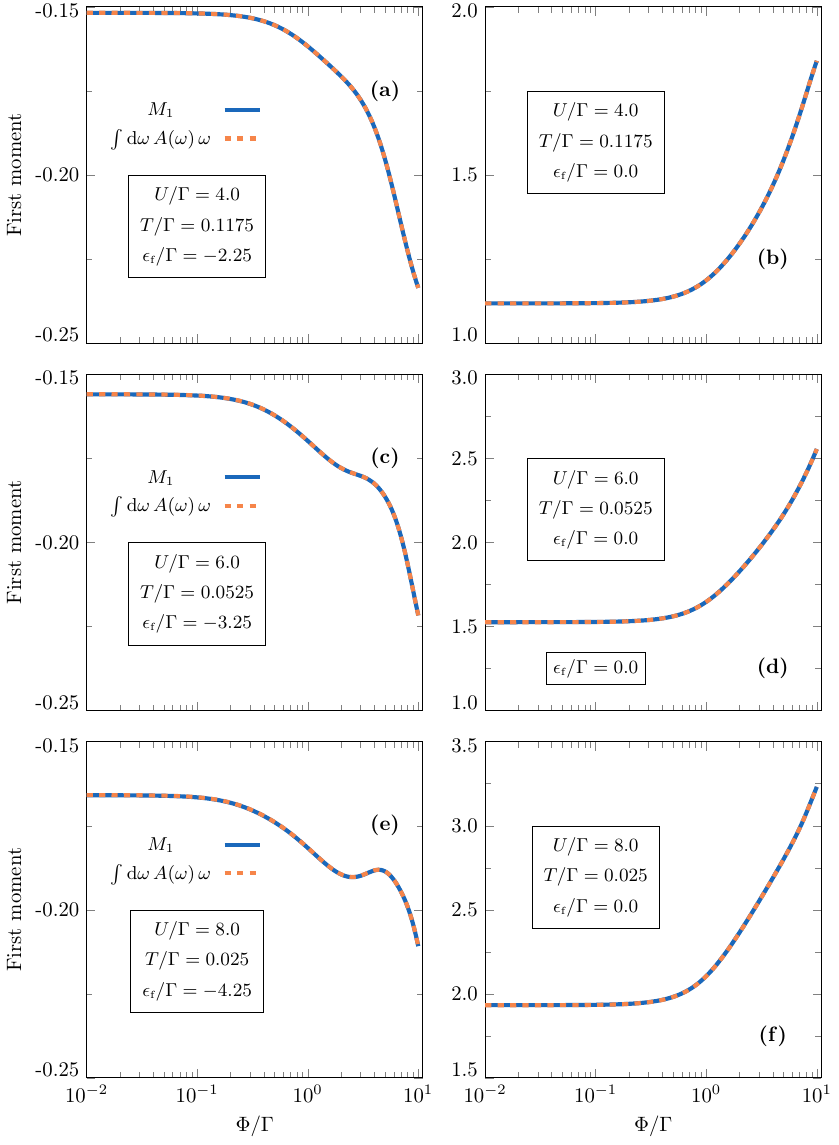}
\caption{First-moment spectral sum rule check, Eq.~\eqref{eq:first_moment}, as a function of bias $\Phi$ for the setups discussed in Fig.~\ref{fig:fig3}. The upper row shows the results for $U=4\Gamma$, the central row for $U=6\Gamma$ and the lower one for $U=8\Gamma$. Temperatures and onsite energies are reported on the plots.}
\label{fig:first_moment}
\end{figure}

\begin{figure}[t]
\includegraphics[width=\linewidth]{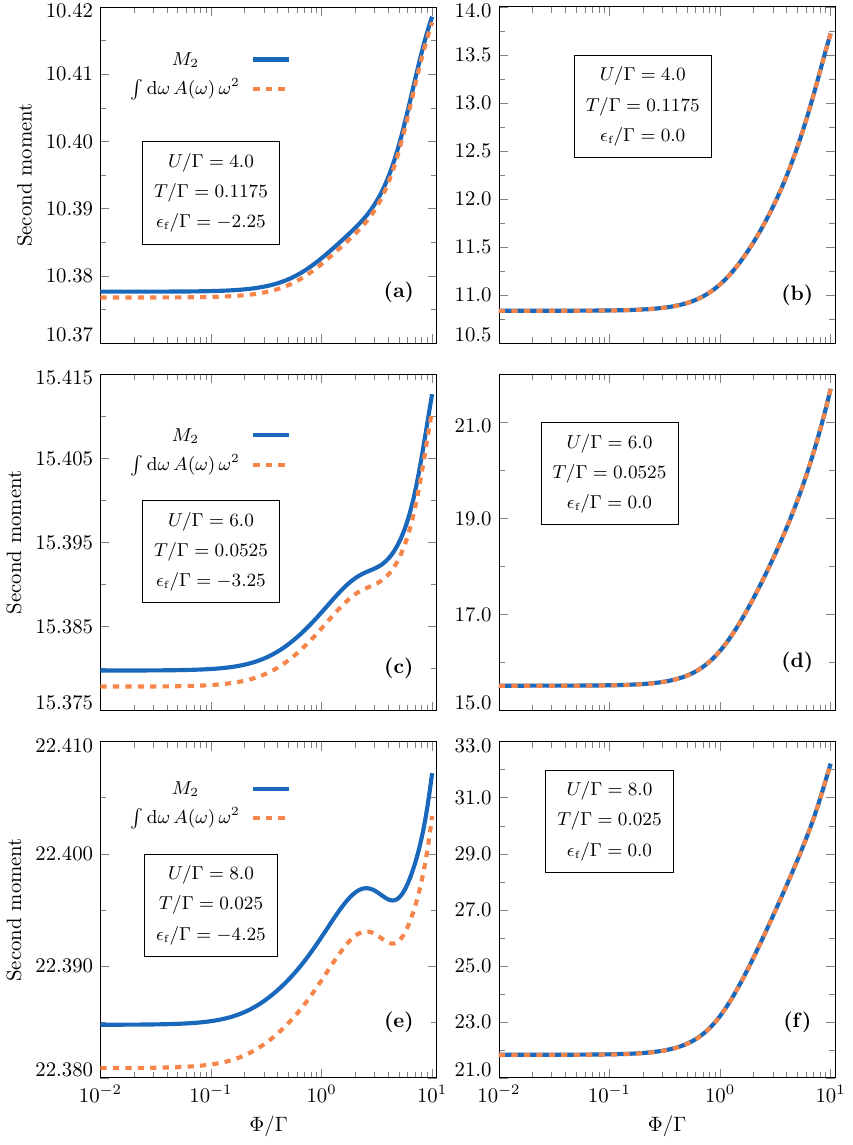}
\caption{Second-moment spectral sum rule check, Eq.~\eqref{eq:second_moment}, as a function of bias $\Phi$ for the setups discussed in Fig.~\ref{fig:fig3}. The upper row shows the results for $U=4\Gamma$, the central row for $U=6\Gamma$ and the lower one for $U=8\Gamma$. Temperatures and onsite energies are reported on the plots.}
\label{fig:second_moment}
\end{figure}

Because these identities involve only the retarded self-energy and Green's function together with the self-consistently determined occupation $n$ [Eq.~\eqref{eq:occupation_num}], they provide an AMEA-independent check of our nonequilibrium KK-IPT solution, complementary to the double-occupancy comparison above. Figures~\ref{fig:first_moment} and \ref{fig:second_moment} show Eqs.~\eqref{eq:first_moment} and \eqref{eq:second_moment} evaluated both from the left and right hand sides of the expressions, for the same setups as Fig.~\ref{fig:fig3}. Across the entire dataset, i.e. all $U$, $T$, fillings and biases considered in this work, the largest relative deviation found is $0.03\%$ for both the first and the second moment, i.e.\ both sum rules are satisfied to better than $0.05\%$ throughout, including the low-temperature, low-bias regime where a controlled AMEA benchmark is not available.

\begin{acknowledgments}
This research was funded by the Austrian Science Fund (FWF) [Grant DOI:10.55776/P33165], and by NaWi Graz. Results have been obtained using the A-Cluster at TU Graz as well as the Austrian Scientific Computing (ASC) infrastructure.
\end{acknowledgments}

\section*{Data Availability}
The data supporting the findings of this study is openly available in the TU Graz Repository at \url{https://doi.org/10.3217/xz6v9-9jp08}. The code used to generate the results discussed in this paper is open source and can be found at \url{https://github.com/mazzocch/neq-kk-ipt-solver}.

\bibliography{references_database,my_refs}

\end{document}